\documentclass[%
 aip,pof,
 amsmath,amssymb,
 reprint,%
]{revtex4-1}

\usepackage{graphicx}
\usepackage{dcolumn}
\usepackage{bm}
\usepackage[mathlines]{lineno}

\usepackage[utf8]{inputenc}
\usepackage[T1]{fontenc}
\usepackage{mathptmx}
\usepackage{etoolbox}
\usepackage{siunitx}
\usepackage{xcolor}
\makeatletter
\def\@email#1#2{%
 \endgroup
 \patchcmd{\titleblock@produce}
  {\frontmatter@RRAPformat}
  {\frontmatter@RRAPformat{\produce@RRAP{*#1\href{mailto:#2}{#2}}}\frontmatter@RRAPformat}
  {}{}
}%
\makeatother

\begin{document}

\preprint{AIP/123-QED}

\title[Influence of the porosity pattern on the aerodynamics of a square plate]{Influence of the porosity pattern on the aerodynamics of a square plate}
\author{A. Gayout}
 \altaffiliation[Also at ]{
Biomimetic Group, Energy and Sustainability Research Institute Groningen, Faculty of Science and Engineering, University of Groningen, 9747 AG Groningen, The Netherlands}
 \email{a.m.m.gayout@rug.nl}
\author{M. Bourgoin}
\author{N. Plihon}
\affiliation{Univ Lyon, ENS de Lyon, CNRS, Laboratoire de Physique, F-69342 Lyon, France}%

\date{\today}

\begin{abstract}
The evolution of the normal aerodynamic coefficient of 19 configurations of square plates with various porosity patterns, ranging from solid plate to homogeneous porous plate, is experimentally characterized. The variation of the porosity pattern is obtained by partially covering the holes of a commercial fly-swatter using adhesive tape. Evolution of the normal aerodynamic coefficient is assessed from the measurement of the angular position of the porous plate, placed as a freely rotating pendulum swept by a flow in a wind tunnel. These angular measurements are also supported by PIV measurements of the structure of the wake. We show that the porosity pattern determines whether or not an abrupt stall occurs. In particular, the details of the porosity pattern on the edges of the plate are decisive for the existence of abrupt stall.   
\end{abstract}

\maketitle

\section{Introduction}

The omnipresence of porous structures in Nature and in technological applications induces complexity in a wide range of physical problems. In the context of technological applications, the water flow through nets and the drag exerted on net structures is crucial for aquaculture~\cite{Klebert2013}. The use of porous structures and fences has long been proposed as a means of controlling the characteristics of flows, with a number of important applications in aerodynamics or civil engineering~\cite{Laws1978,Lee1999}.
The development of fog harvesters for water supply in arid regions requires a fine understanding of flows in the vicinity of nets~\cite{Regalado2016,Moncuquet2022}. At smaller scales, the efficiency of respiratory masks to reduce the propagation of airborn viruses rely on reducing the amount and distance of aerosols spread following the  propagation of multiphase flows through finely meshed masks~\cite{Mittal2020,Bourouiba2021}. In the context of wind-dispersed plant seed, the flight of the dandelion involves a porous structure made of a bundle of bristles. It was recently demonstrated that the aerodynamic drag was maximized thanks to a specific structure of the wake, namely a separated of a vortex ring~\cite{Cummins2018,Ledda2019}. Bristled wing are also widespread in Nature and were shown to increase lift at small Reynolds numbers \citep{Sunada2002,Kolomenskiy2020}. 

Fundamental aerodynamics studies of porous materials date back to the pioneering work of G.I. Taylor~\cite{Taylor1944_1,Taylor1944_2}, and largely focused on the influence of porosity on pressure drop and on drag at normal incidence.
Castro investigated the influence of the porosity fraction of perforated plates with centimetric-holes and showed a drag decrease as the porosity fraction decreases and the absence of vortex shedding for porous fractions 
above $\sim 0.2$. Several experimental studies then focused, for instance, on the interaction of periodically arranged jets emerging from porous screens~\cite{Villermaux1994} or the shape of the perforated obstacle~\cite{Steiros2020}. The modelling of the flow behind porous screen at normal incidence also attracted attention~\cite{Koo1973,Steiros2018}. Surprisingly, the effect of the angle of attack on the aerodynamic coefficients and the flow features of porous screen has only recently been studied experimentally and theoretically~\cite{Marchand2023}, though it was shown that porosity strongly influences the trajectory of permeable or porous disks~\cite{Vincent2016}, or its stability~\cite{Vagnoli2022}. In this article, we extend these previous work and study experimentally the aerodynamic coefficients of porous plates with inhomogeneous porosity pattern, and span a wide range of angles of attack. Our strategy was to systematically vary the porosity pattern by partially covering a porous plate with a large numbers of holes, and we chose an ideal object widely spread for this study: a fly-swatter.

\begin{figure}[t]
\centering
\includegraphics[width=0.95\linewidth]{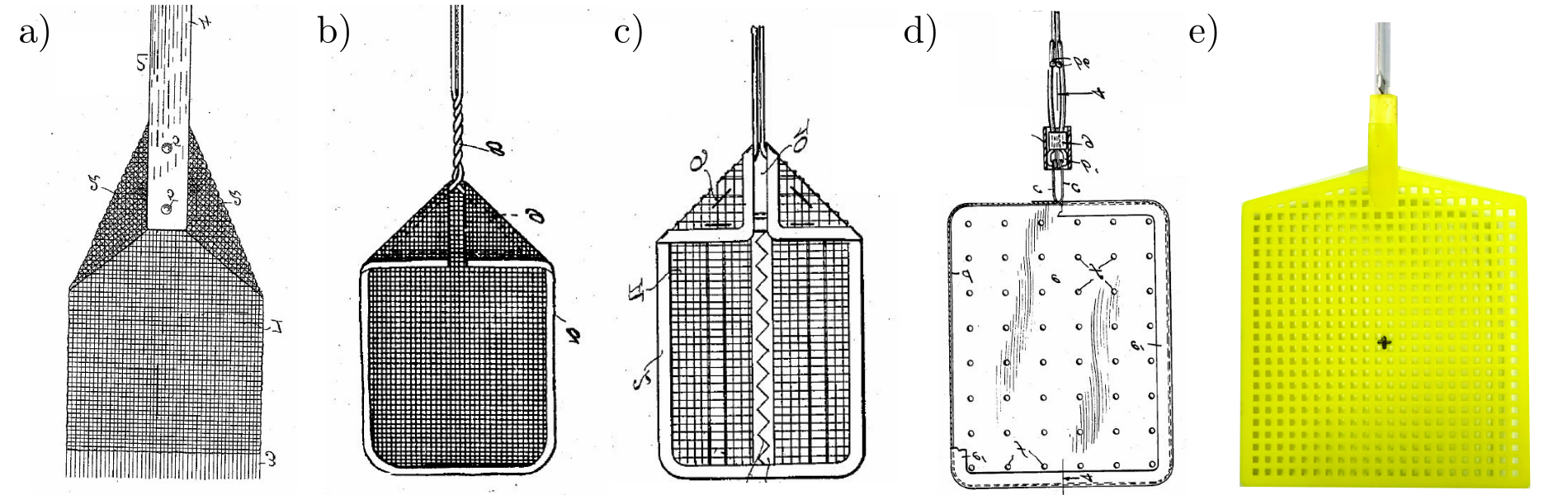}
\caption{a) Fly-killer in 1900 from \citep{Montgomery1900}. b) Fly-swatter in 1927 from \citep{Gatch1927}. c) Fly-swatter in 1938 from \citep{Brownson1938}. d) Fly-swatter in 1939 from \citep{Baker1939}. e) Fly-swatter used in this article .}\label{fig:designswatter}
\end{figure}

Patented in 1900 by Montgomery, the first modern fly-swatter, the ``Fly-Killer'', was composed of a rectangular wired net \citep{Montgomery1900}. The use of wire-netting was introduced for durability and elasticity but no reference on the aerodynamic advantage of such netting is mentioned in the patent. In the later patents of Gatch in 1927 \citep{Gatch1927} and Brownson in 1938 \citep{Brownson1938}, improvements on the fly-swatter mostly cover the handle, to facilitate the killing motion. In 1939, Baker patented a different kind of fly-swatter, made of a rubber surface with a few holes, which is supposed to act as a pocket to trap the fly with the elastic recoil from the surface killing the fly without crushing it \citep{Baker1939}. This change was motivated by avoiding property damage and traces when battling with and killing a fly. Here again however, no aerodynamic considerations are presented in the patents and it seems that holes were added empirically, to either reduce costs or increase elasticity. Sketches of the above-mentioned fly-swatter are reproduced in Fig.~\ref{fig:designswatter}.a-d), together with a photograph of the modern plastic model used for the present investigation.

The evolution of the normal aerodynamic coefficient (see next section) when varying the angle of attack and the porosity was obtained placing the fly-swatter at the bottom of a freely rotating rod, acting as a pendulum, and placed in a windtunnel, following a setting previous investigated~\cite{Obligado2013,Gayout2021}. As the wind speed increases, the angular position evolves and is set by the torque balance~\cite{Obligado2013,Gayout2021}. When an abrupt stall occurs, the result is a discontinuity in angular position between a drag-dominated branch and a lift-dominated branch, possibly leading to bistability~\cite{Obligado2013,Gayout2021}. The transitions between the two branches were shown to be controlled by rare aerodynamics events~\cite{Gayout2021}, making this aerodynamic pendulum one of the simplest experimental configuration to study rare-event statistics.

This article is organized as follows: Sec.~\ref{sec:experimental} describes the experimental setup. The results are then presented and discussed in Sec.~\ref{sec:results}, with a focus on the differences between a solid and a porous plate in Sec.~\ref{sec:solidporous}, and the investigation of 19 different porosity pattern in Sec.~\ref{sec:concentric} and \ref{sec:edge}. Conclusions are then drawn in Sec.~\ref{sec:conclusion}.

\section{Experimental setup}\label{sec:experimental}

The influence of the porosity pattern on the aerodynamic coefficients has been experimentally evaluated using the plastic fly-swatter shown in Fig.~\ref{fig:designswatter}.e). The plastic fly-swatter consists in a square of size $a=\SI{10}{\centi\meter}$ with square holes of size \SI{2.4}{\milli\meter} equally spaced at a distance of \SI{1.8}{\milli\meter}. Each row and column of the square section contains $22$ holes, for a total of  $484$ holes. A small triangular shape at the top of the fly-swatter, with 26 holes on each side, connects the square section to the fly-swatter holder. In this article, we will focus on the influence of the porosity pattern of the square section on the aerodynamic coefficients of the system, and the holes of the upper triangular part will be left open. The porosity pattern is modified by adding adhesive vinyl tape to block specific holes (see Sec.~\ref{sec:concentric} for details). The vinyl tape was placed so that only rows or columns of 2-hole width were sealed by one piece of tape, and no tape peeling was observed over the various experiments. The maximum  porosity, given as the ratio between the surface of the 484 holes and the surface~of~the~whole~square is $\simeq 30\%$. Of the $2^{484}$ possible configurations for the partial covering of the fly-swatter, we decided to select only $19$ with left-right symmetry and to focus on the onset of stall when the porosity pattern is modified. For $11$ of them, due to a slight curvature of the fly-swatter, two sets of measurements were carried out, one with the curvature facing upstream and the other downstream, as will be discussed in Sec.~\ref{sec:curv}.

\begin{figure}[ht]
\centering
\includegraphics[width=0.65\linewidth]{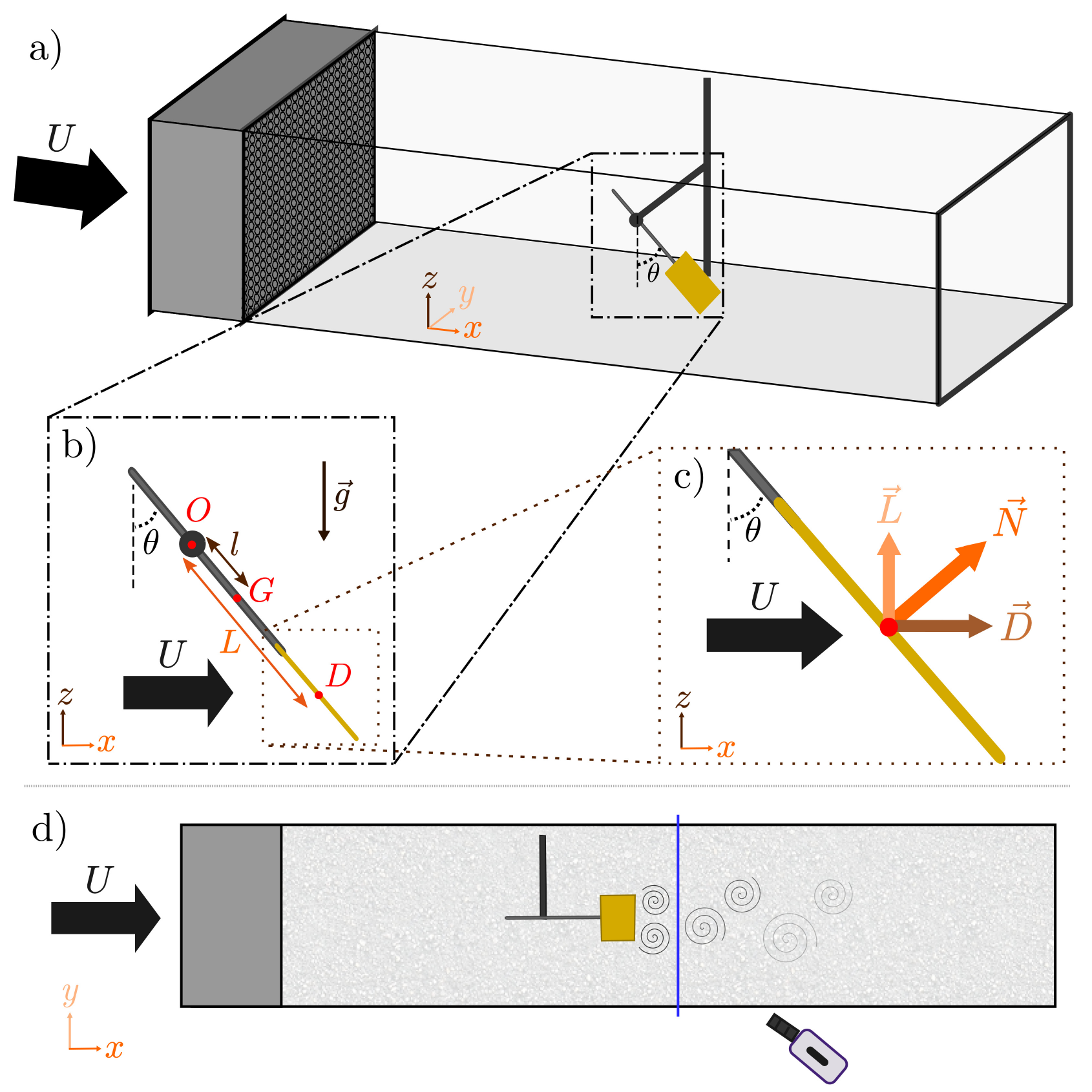}
\caption{a) Schematic view of the wind tunnel with the fly-swatter. b) Details of the pendular attachment of the fly-swatter. c) Definition of the aerodynamic forces. d) PIV setup overview. The camera focus is redressed by a Scheimpflug apparatus. }\label{fig:windtunnel}
\end{figure}

The aerodynamic coefficients of the plates with various porosity patterns are assessed from measurements in a wind tunnel, as sketched in Fig.~\ref{fig:windtunnel}, following the protocol detailed elsewhere~\cite{Obligado2013, Gayout2021, Gayout2023}, and recalled here. The fly-swatter, placed facing the flow, is attached at the extremity of a rod free to rotate around point $O$. Frictionless rotation of this pendulum is obtained using an air bushing (OAVTB16i04 from OAV Labs), and the angular position $\theta$  with respect to the vertical is recorded by a contact-less rotary encoder with minimal friction (DS-
25, 17-bit digital encoder from Netzer). The distance between the pivot point and the center of the fly-swatter is $L=\SI{17.3}{\centi\meter}$. The distance $l$ between point $O$ and center of mass G of the pendulum is computed for each configuration, knowing the mass of the rod and of the fly-swatter and measuring the mass of the added vinyl tape strips. 
The flow impinging the porous square exerts an aerodynamic torque $\Gamma_\mathrm{aero}$ at point $O$, due to both lift and drag, and the equation of motion reads 
 $J\ddot{\theta} = - m g l \sin{\theta} + \Gamma_\mathrm{aero},$ with $J$ the moment of inertia of the pendulum.
The aerodynamic forces acting on the fly-swatter (see Fig.~\ref{fig:windtunnel}c) are the drag force $\mathbf{D}=  1/2 \rho L U^2 a^2 C_D(\theta) \mathbf{e_x}$ and the lift force $\mathbf{L}=  1/2 \rho L U^2 a^2 C_L(\theta) \mathbf{e_z}$, with $\rho$ the air density, $a$ the side of the square plate and where the drag coefficient $C_D$ and the lift coefficient $C_L$ depend on $\theta$~\cite{Flachsbart1932}. The total aerodynamic torque is expressed as $ 1/2 \rho L U^2 a^2 C_N(\theta)$, where the normal coefficient $C_N$ is defined
as $C_D \cos{\theta} +  C_L \sin{\theta}$~\cite{Obligado2013}. In this article, we will focus on steady state regimes, for which the torque induced by the weight of the
pendulum balances the aerodynamic torque:
\begin{equation}\label{eq:pendtap}
 mgl\sin(\theta)=\frac{1}{2}\rho U^2 a^2 L C_N(\theta)
\end{equation}
We made the choice of ignoring the covering fraction for the area $a^2$ of the fly-swatter used to define the aerodynamic coefficients, as we expect it to not be a simple proportionality factor due to the aerodynamic coupling of holes in the array. As mentionned above, the mass $m$ has been measured for each configuration, due to the vinyl tape adding up to \SI{2}{\gram} to the fly-swatter when fully covered, and $l$ was computed accordingly.
For each configuration, the mean angle $\theta$ is measured over at least \SI{15}{\second} for each flow velocity, ensuring statistical convergence of the mean-value, and the normal aerodynamic coefficient $C_N(\theta)$ is computed from Eq.~\ref{eq:pendtap}.

In addition to the recording of the angular position, Particle
Image Velocimetry (PIV) has been implemented in the wind tunnel to enable flow visualization in the wake of the fixed fly-swatter. This flow visualization is done in the transverse $(y, z)$ plane. This choice was in particular motivated by the tri-dimensionality of the wake for pendulums of aspect ratio close to 1~\cite{Gao2018}. The flow structure is obtained from smoke particle imaging. Particles are illuminated from a laser sheet produced by a \SI{5}{\watt} blue diode laser through a Powell lens with a \SI{30}{\degree} fan angle, and imaged using a high-speed camera (Phantom v26.40) at a resolution of 2048 by 1952 pixels and a \SI{100}{\milli\meter} lens through a Scheimpflug adaptor. As we chose to visualize the transverse $(v,w)$ flow in the $(y,z)$ plane , the particles only remain in the sheet for a short time, which imposes constraints on the flow velocity (set to \SI{1.7}{\meter\per\second}), the thickness of the laser sheet (\SI{4}{\mm} was observed to be an optimal choice) and the framerate (set to \SI{2000}{fps}). The PIV algorithm uses the open software UVMAT~\cite{bib:uvmat}.



\section{Results}\label{sec:results}
\subsection{Solid versus homogeneous porous square}\label{sec:solidporous}

Let us first focus on the main differences between a solid square (i.e. all holes have been covered by vinyl tape) and a homogeneous porous square (i.e. the original fly-swatter geometry), before investigating more complex patterns in the following subsections.

\begin{figure*}[ht!]
\centering
\includegraphics[width=0.75\linewidth]{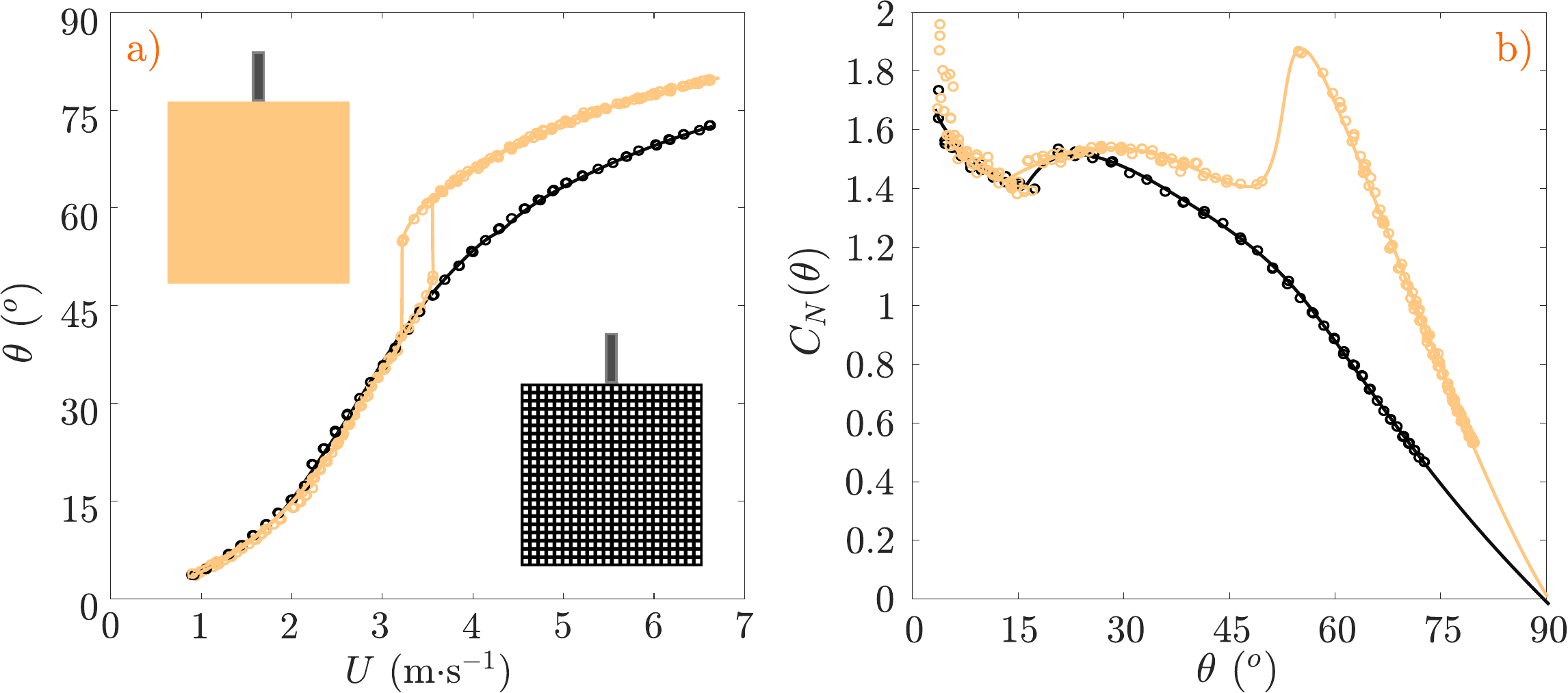}
\caption{Aerodynamic response of the fly-swatter (black) compared to a square plate (beige): a) evolution of $\theta$ as a function of $U$, b) $C_N$ coefficient computed from a) and Eq. \ref{eq:pendtap}. The curves are spline interpolations and thus an eye guide.}\label{fig:Cnfullempty}
\end{figure*}

Figure~\ref{fig:Cnfullempty}.a) shows the evolution of the angular position $\theta$ as a function of the wind velocity $U$ for the solid square (orange) an the porous square (black). The evolution of the normal aerodynamic coefficient $C_N(\theta)$ as a function of $\theta$,  computed from Eq.~\ref{eq:pendtap}, is shown in Fig.~\ref{fig:Cnfullempty}.b). We note that $\theta(U)$  exhibits abrupt transitions and bistability for the solid plate, which can be readily understood from the evolution 
of the steady-state aerodynamic coefficients (see detailed discussions in Ref.~\cite{Obligado2013,Gayout2021}). For velocities below the bistable region (and angles below \SI{50}{\degree}), the dominant aerodynamic force is the drag force, and corresponds to a nearly constant $C_N$ value.  For velocities above the bistable region (and angles above \SI{53}{\degree}), the dominant aerodynamic force is the lift force. The stall angle corresponds to a dramatic decrease of $C_N$ with decreasing angle, namely at around \SI{52}{\degree} for the solid square, and translates in abrupt transition in the evolution of $\theta$ as a function of $U$. We stress that this stall angle is in agreement with values observed for flat square by Eiffel ($\theta_{stall}\simeq\SI{51}{\degree}$)~\cite{Eiffel1910} and Flachsbart ($\theta_{stall}\simeq\SI{50}{\degree}$)~\cite{Flachsbart1932}. Remarkably, the transition is smoother for the homogeneous porous square, and there is no abrupt stall angle. We note two noteworthy features at low angles, for both configurations. The first one is a strong increase of $C_N$ as $\theta$ is decreased below a \SI{10}{degrees}, and is discussed further in Sec.~\ref{sec:curv}. The second one is the bump observed on $C_N$ between  \SI{15}{\degree} and \SI{18}{\degree}, which is attributed to the presence of stall on the holding rod. In the remaining of this article (apart from Sec.~\ref{sec:curv}), we will thus ignore the evolution of the $C_N$ coefficient below \SI{18}{\degree}, and we will focus on the conditions for which an abrupt stall is observed when varying the porosity pattern (data below \SI{18}{\degree} will be systematically shown as lighter symbols in the remaining of this article).

\begin{figure*}[ht]
    \centering
    \includegraphics[width=\linewidth]{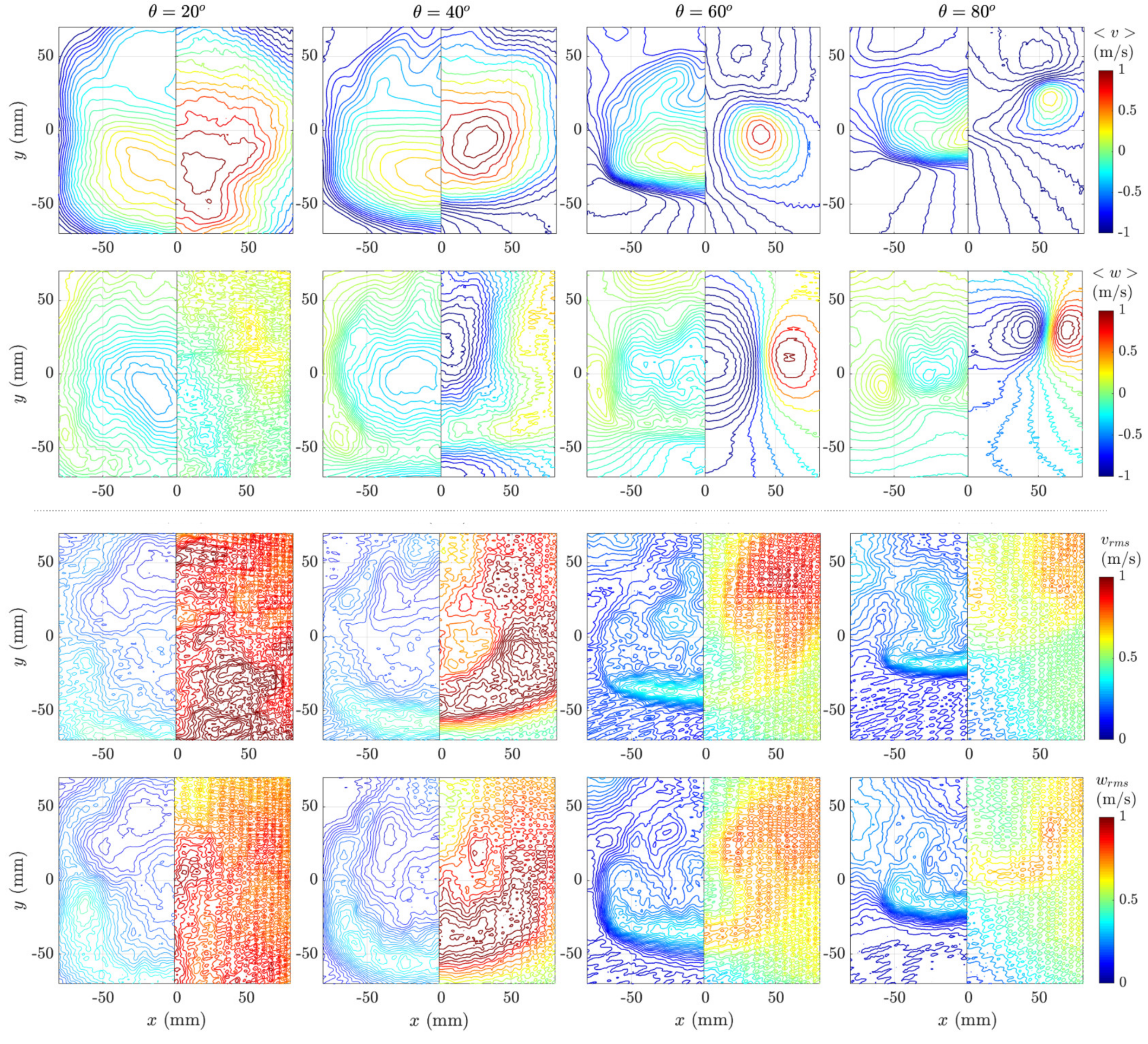}
    \caption{Wake structure behind the hollow fly-swatter (left side) and fully covered fly-swatter (right side) for \SI{20}{\degree}, \SI{40}{\degree}, \SI{60}{\degree} and \SI{80}{\degree}. Upper part: mean velocity -- transverse mean $<v>$ (top) and vertical mean $<w>$ (bottom). Lower part: velocity fluctuations -- transverse fluctuations $v_{rms}$ (top) and vertical fluctuations $w_{rms}$ (bottom). For the hollow fly-swatter, the wake structure is consists primarily of a trailing-edge vortex and does not change with the angle, apart from a vertical shrinking due to the vertical projection of the fly-swatter diminishing. For the fully covered fly-swatter, the intensity of fluctuations is much higher than the hollow fly-swatter and the structures are more difficult to identify. Above \SI{60}{\degree}, a trailing vortex is visible in the upper region of the wake.}
    \label{fig:piv_tap}
\end{figure*}

The PIV measurements of the transverse components of the velocity field in the wake, 10 cm downstream of the center of the fly-swatter are shown in Fig.~\ref{fig:piv_tap}. For these measurements, the angle $\theta$ was set to four different constant values (namely \SI{20}{\degree}, \SI{40}{\degree}, \SI{60}{\degree} and \SI{80}{\degree}), with no free rotation allowed around point $O$, i.e. this is not a pendulum configuration. In each panel, the flow of the porous plate (the original fly-swatter) is displayed on the left half while the flow of the solid plate (fully covered fly-swatter) is shown on the right half. The two upper rows show the time-averaged structure of the wake, and the two lower rows the amplitude of the fluctuations. The PIV measurements are shown here with the purpose of illustrating the differences in flow features between the wake created by solid and porous plates, but no deep quantitative analysis is presented.
We note strong differences between both configurations. In the case of the solid plate, there is a clear difference below and above the stall angle ($\simeq \SI{50}{\degree}$). Strong trailing vortices (also known as wingtip vortices) with a large down-wash at the center are observed in the mean flow above the stall angle (\textit{i.e.} \SI{60}{\degree} and \SI{80}{\degree}). In contrast, the structure of the mean-flow is similar for all inclination angle for the porous plate; a feature shared with the fluctuating part. The maximal level of fluctuations of the solid plate is about three times that of the porous one. The influence of the stall angle for the solid plate is also evident, with very localized fluctuations are observed around the vortices above the stall angle.

\subsection{Reducing the fraction of porous surface by concentric holes covering}\label{sec:concentric}

\begin{figure*}[ht]
\centering
\includegraphics[width=0.95\linewidth]{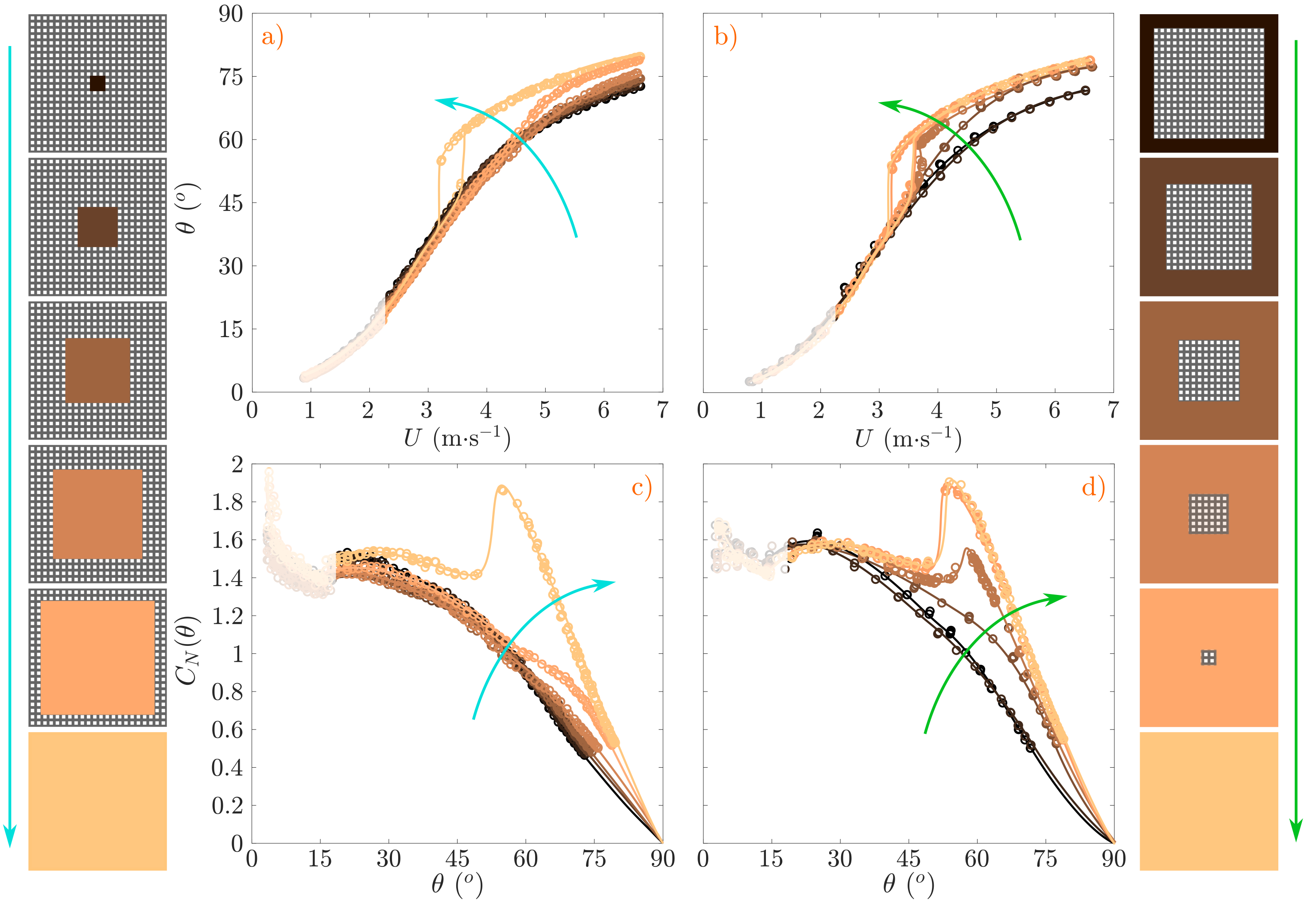}
\caption{Influence of the concentric sealing of holes on the angular equilibrium positions and the $C_N$ coefficient. Top (a,b):  evolution of $\theta$ as a function of $U$ for different hole-covering configurations. Bottom (c,d): associated $C_N$ coefficient computed using Eq. \ref{eq:pendtap}. Left (a,c): concentric covering starting from the center towards the edges. Right (b,d): concentric covering from the edges towards the center. Color codes for the configuration: the lighter the color, the more holes sealed. The curves are spline interpolations and thus an eye guide.}\label{fig:concentric}
\end{figure*}

The porosity pattern was first modified by concentric holes covering in two different ways: either from the center towards the edges or vice-versa. Six different configurations, for each concentric direction, are investigated, with five of them being partially porous and obtained by the addition of vinyl tape strips covering a two-hole wide band, as indicated in the side sketches of  Fig.~\ref{fig:concentric}. When the fraction of porous surface is reduced from the center (left half in Fig.~\ref{fig:concentric}), no significant changes are observed on the evolution of $\theta$ with the velocity $U$, until covering the last two lines of holes on the edges. In particular, bistability is only observed for the fully taped configuration (see Fig. \ref{fig:concentric}.a). This observation is in contrast with the behavior observed when the porous fraction is reduced when covering the holes from the edges towards the center (right half in Fig.~\ref{fig:concentric}). Bistable regimes are observed for several configurations, as soon as the central porous square is smaller than a  $6 \times 6$-hole square (see Fig.~\ref{fig:concentric}.b). Almost no difference is observed between the solid case and the case when a $2 \times 2$-hole square is left open at the center. For bistable configurations, the evolution of the $C_N$ coefficient with $\theta$ displays a sharp increase at the stall angle (see Fig.~\ref{fig:concentric}.c and d).


These results highlight the importance of the porosity of the edges for the existence of a sharp stall: the presence of holes on the edges of the plate prevent the occurrence of a sharp stall. On the contrary, stall can be readily observed in the presence of a significant porous fraction at the center of the plate.

\subsection{Triggering stall from the edge porosity 
pattern}\label{sec:edge}

\begin{figure*}[htp]
\centering
\includegraphics[width=0.8\linewidth]{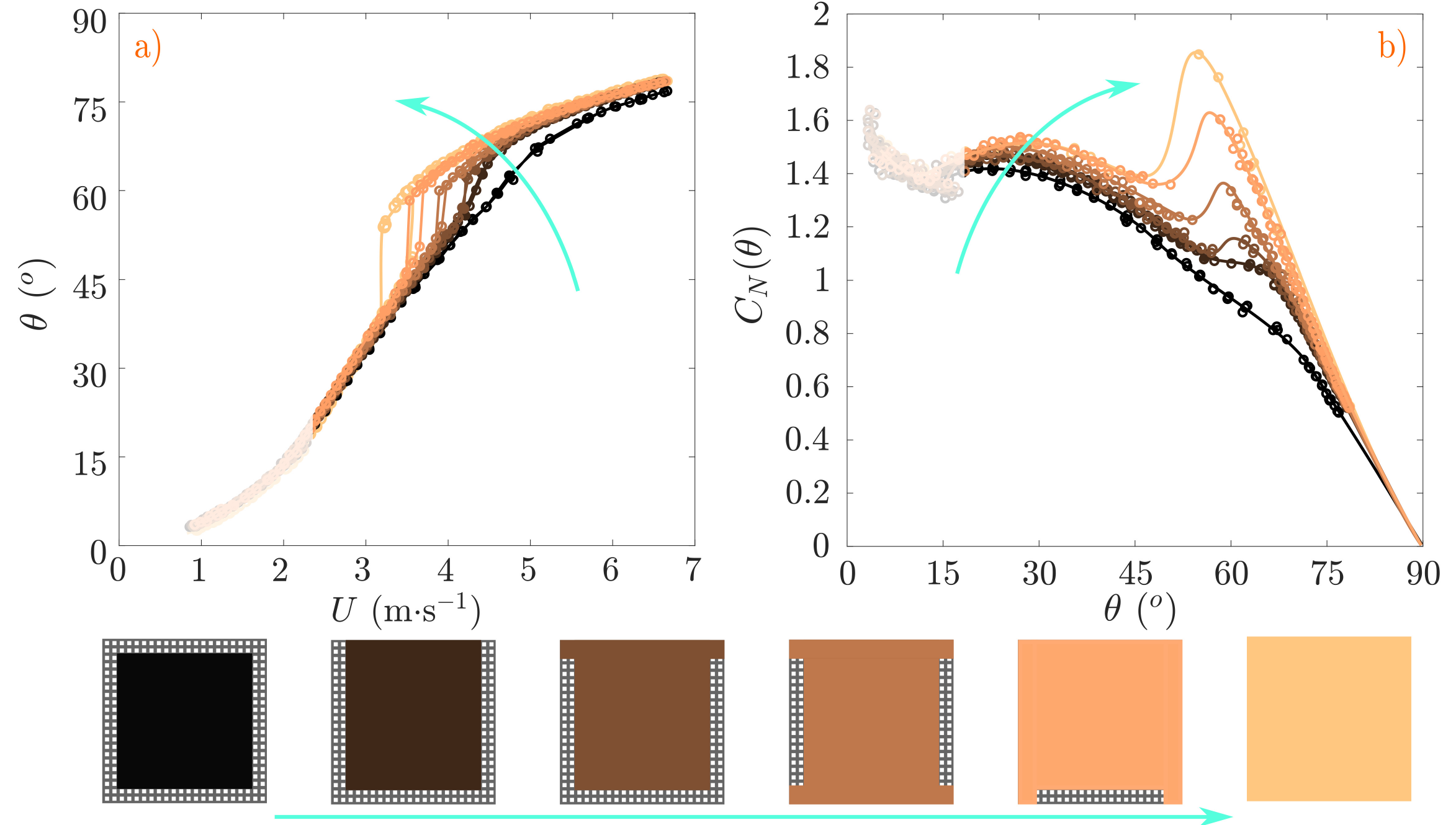}
\caption{Triggering  bistability and sharp stall by partially covering the outer rows of holes on the fly-swatter. Color codes for the configuration, described at the bottom. a) Evolution of $\theta$ as a function of $U$. b) $C_N$ coefficient computed from a) and Eq. \ref{eq:pendtap}. The curves are spline interpolations and thus an eye guide.}\label{fig:partialbistability}
\end{figure*}

\begin{figure*}[htp]
\centering
\includegraphics[width=0.8\linewidth]{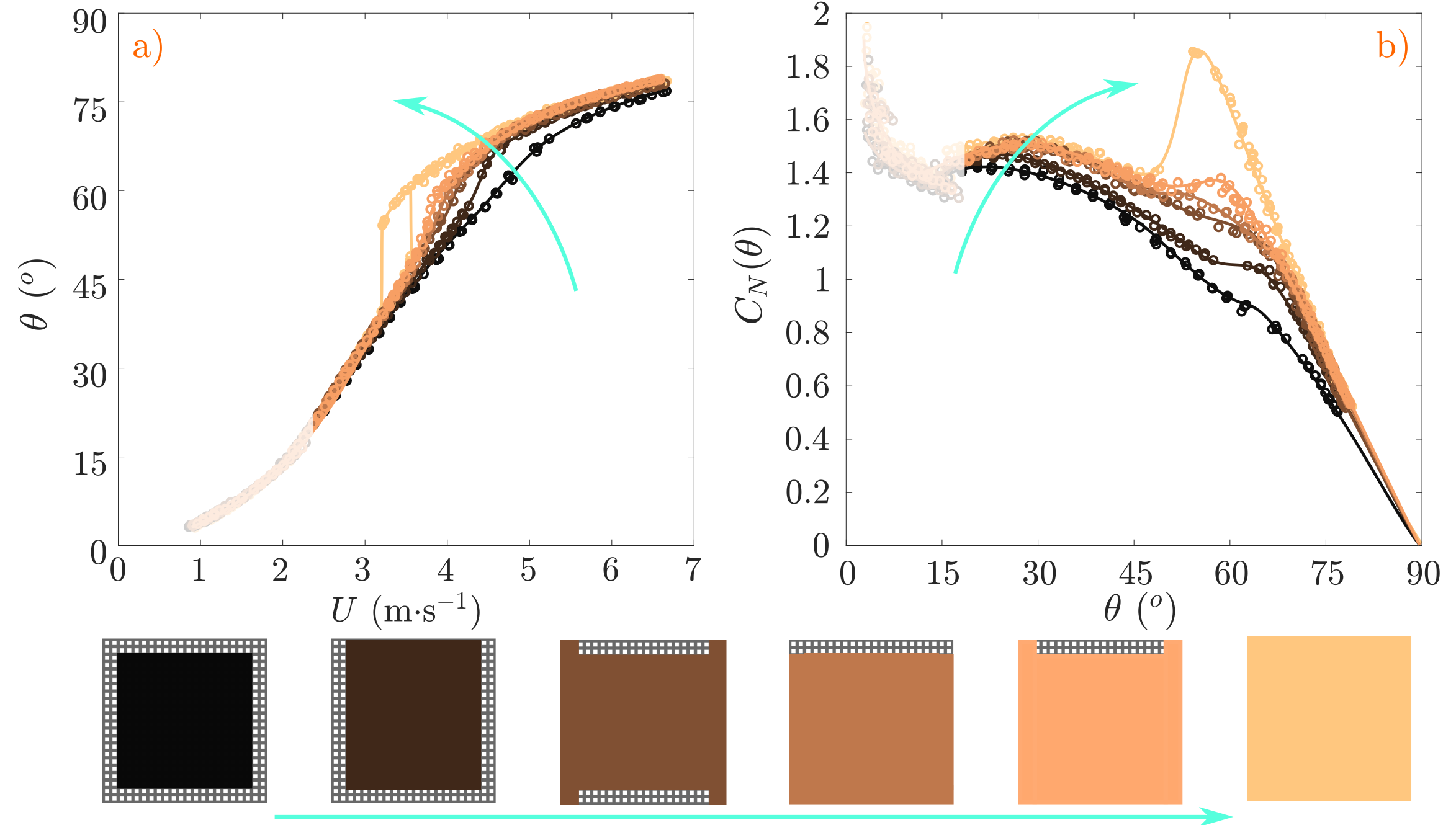}
\caption{Partially covering the outer rows of holes on the fly-swatter without achieving bistability. Color codes for the configuration, described at the bottom.  a) Evolution of $\theta$ as a function of $U$. b) $C_N$ coefficient computed from a) and Eq. \ref{eq:pendtap}. The curves are spline interpolations and thus an eye guide.}\label{fig:partialnobistability}
\end{figure*}

Section \ref{sec:concentric} highlighted that bistability on the evolution of $\theta(U)$ and abrupt stall only appear when the last two lines of holes at the periphery of the plate are covered, when all the inner holes are covered. In order to better understand influence of edge porosity on the emergence of abrupt stall, seven configurations with a partially covered periphery were tested. Note that some of these configurations break the top-down symmetry, and that, following the convention of Fig.~\ref{fig:Cnfullempty}, the holding rod of the various configuration is placed at the top part of the sketches. Fig.~\ref{fig:partialbistability} focuses on configurations for which the bistability is observed as the fraction of porous surface of the plate is decreased. Configurations, that, despite a similar porous fraction, present no bistability, are shown in Fig.~\ref{fig:partialnobistability}.

The two leftmost configurations shown in Fig. \ref{fig:partialbistability} do not present any bistability, whereas bistability develops for the four rightmost configurations. When bistability is observed, the range of bistable positions increases with the covering fraction, \textit{i.e.} from left to right. The bistability thus first arises, when in addition to the center of the fly-swatter, the top two rows are fully covered with tape. Covering the upper 4-holes corners on each side seems critical for the onset of bistability, since, when they remain uncovered (as in the second leftmost configuration), no bistability was observed.

The signature of bistable regimes is also readily observed on the evolution of $C_N(\theta)$ (see Fig.~\ref{fig:partialbistability}.b). A local increase of $C_N(\theta)$ is observed for angles between 50 and \SI{70}{\degree}, seemingly correlated with the stall angle, that separates the lift and the drag branches. More precisely, we note that covering the two upper corners only induces a bump on the  $C_N$ coefficient between \SI{57}{\degree} and \SI{65}{\degree} (third leftmost configuration). This modest increase leads to an inflection point on the $C_N$ coefficient, which is nonetheless sufficient to induce a sharp stall. As the porous fraction decreases further, the stall angle occurs for lower $\theta$ angles, decreasing from from \SI{62}{\degree} to \SI{52}{\degree}. The lift-dominated regime thus occurs over a larger range of angles $\theta$ (we recall that $\theta= \pi/2 - \alpha$, with $\alpha$ the angle of attack). Consequently, since bistability is observed around the stall angle, it occurs for angles that decrease as the porous fraction decreases. The range of velocity for which bistability occurs also increases as the porous fraction decreases. Surprisingly though, the span of forbidden angles (which are never explored neither in the lift branch nor in the drag branch) remains almost constant, around \SI{6}{\degree}. Understanding whether this observation might be linked to the structure of the wake would require an extensive dataset of 3D flow measurements, out of reach of the present study.


No bistability, and thus no stall, was observed for any configuration shown in Fig.~\ref{fig:partialnobistability}.a (apart from the solid plate). 
A specificity  of all non-bistable configurations tested thus far is the presence of holes on the upper rows. For all these configuration, the evolution of the $C_N$ coefficient with $\theta$ (see Fig.~\ref{fig:partialnobistability}.b), does not display a strong bump in the range of angles $\theta\in[\SI{45}{\degree},\SI{70}{\degree}]$. We stress here that a weak bump is observed for the fifth leftmost configuration, around $\theta=\SI{58}{\degree}$, but which does not trigger bistable regimes. The values of the $C_N$ coefficient observed form the solid plate on the drag branch ($\theta<\SI{45}{\degree}$) are recovered as soon as the lateral edges of the fly-swatter are covered. 




This detailed study of the influence of the porous pattern at the edges of the plate suggests that a necessary condition for the existence of the bistability is the full covering of the upper rows. 
The configuration that displays bistability with the highest porous fraction is the third leftmost configuration shown in Fig.~\ref{fig:partialbistability}, with only the side and bottom edges left porous. On the other hand, the configuration with the least porosity that present no sharp stall is the second rightmost in Fig.~\ref{fig:partialnobistability}, with only the leading-edge being partially hollowed.

Over the $2^{484}-19$ remaining configurations, \textit{i.e.}$~$ more than $4.9\times10^{145}$ configurations, we cannot rule out that these are not the ones with the respectively highest porous fraction for bistable regimes and lowest porous fraction for the absence of stall. 

\subsection{Curvature effects on $C_N$ coefficient}\label{sec:curv}

Let us now briefly discuss a side effect of using a commercial fly-swatter as the initial porous plate: the effect of the weak natural curvature of the plate on its aerodynamic res\-ponse. This effect was particularly experienced when testing whether the side covered by the tape influences the aerodynamics. The fly-swatter is indeed slightly curved and curvature can have a strong effect on aerodynamic properties, as observed already nearly a century ago by Flachsbart~\citep{Flachsbart1932}.

\begin{figure*}[htp]
\centering
\includegraphics[width=0.9\linewidth]{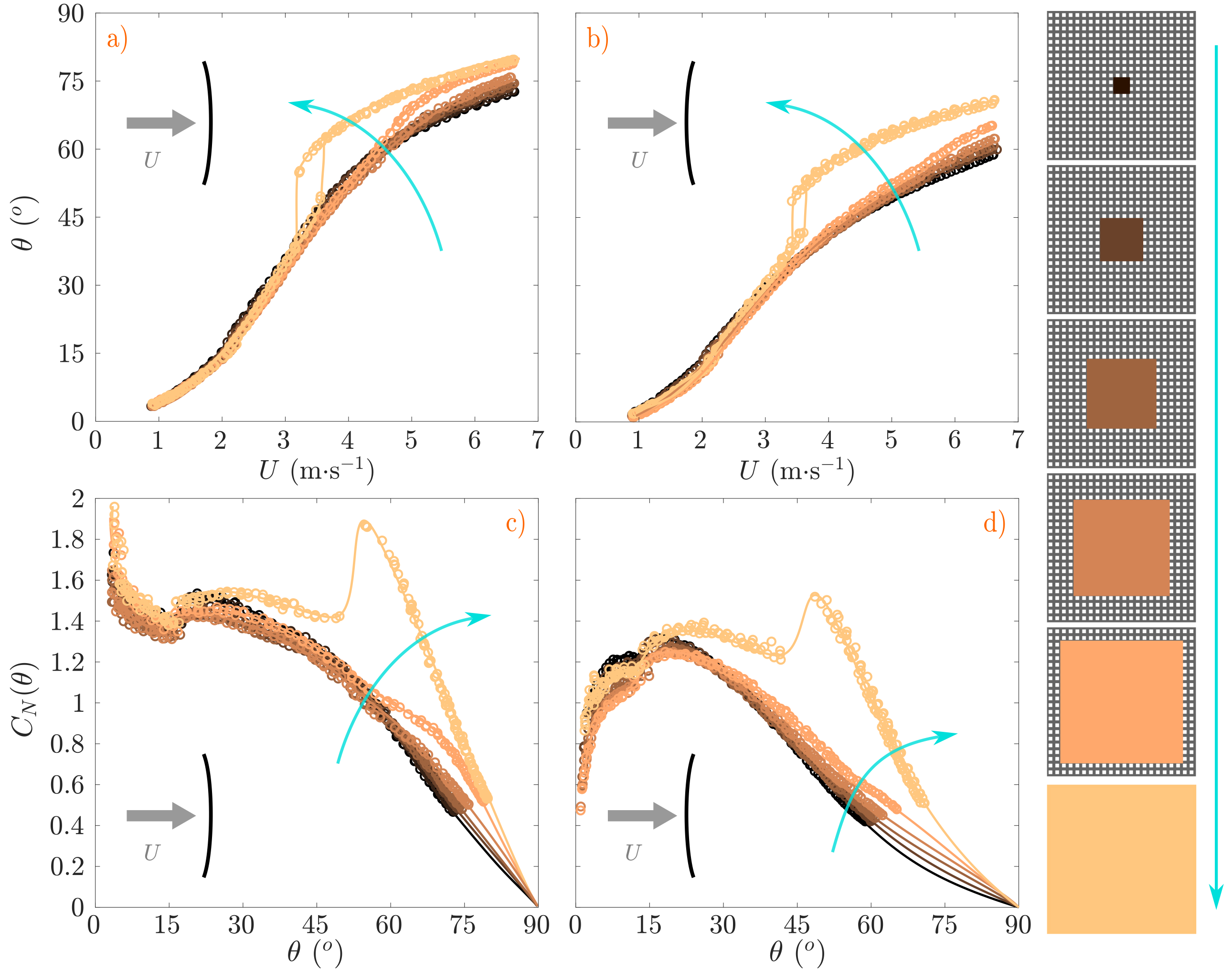}
\caption{Influence of curvature on the aerodynamic response of the partially covered fly-swatter. Top (a,b):  evolution of $\theta$ as a function of $U$ for two curvature configurations with concentric covering starting from the center towards the edges. Bottom (c,d): associated $C_N$ coefficient computed using Eq. \ref{eq:pendtap}. Left (a,c): concave side facing upstream. Right (b,d): convex side facing upstream. Color codes for the configuration, same as in Fig. \ref{fig:concentric}.a and \ref{fig:concentric}.c. The curves are spline interpolations and thus an eye guide.}\label{fig:curvature}
\end{figure*}

Four series of additional experiments were carried out for the concentric covering  configurations, for which the face facing the incoming air flow could be either the concave or convex face, and could be the face covered or not by the adhesive tape. No influence of the side on which tape is glued was observed, while curvature  orientation seems to greatly alter the aerodynamics, as shown in Fig.~\ref{fig:curvature}.


Let us first discuss the influence of curvature on the evolution of $\theta(U)$ for the six configurations shown in Fig.~\ref{fig:curvature}.a) and b). Several  differences are noted : 1) the bistable zone is much narrower when the convex side faces upstream (b), 2) for the same flow velocity, for instance $U=\SI{6}{\meter\per\second}$, the angular position is lower  when the convex side faces upstream.

These observations lead to large differences for the values of the $C_N$ coefficient due to curvature (see Fig.~\ref{fig:curvature}.c and d).
On average, for all configurations and all angles, the $C_N$ coefficient is much higher when the concave side faces upstream. A striking observation is on the sharpness of the stall. For the fully-covered fly-swatter, the stall is indeed much smaller when the convex side faces upstream, with a factor of two for the amplitude of the discontinuity. The difference is also striking when the pendulum is close to the vertical position (\textit{i.e.} for $\theta$ values below \SI{18}{\degree}), for which $C_N$ appears to diverge when the concave side faces upstream (Fig.~\ref{fig:curvature}.c), while it decreases to 0 when the convex side faces upstream (Fig.~\ref{fig:curvature}.d). This feature is observed for all covering configurations, which supports the conjecture that this is an effect of curvature. 

All results presented in Sec.~\ref{sec:solidporous} and \ref{sec:concentric} and \ref{sec:curv} were obtained with the concave side facing the flow (\textit{i.e.} the configuration of Fig.~\ref{fig:curvature}.a and c). Based on the observations for the concentric covering configurations, we expect to observe results similar to those reported in Sec.~\ref{sec:edge} when the convex side faces the flow.

\section{Conclusion}\label{sec:conclusion}

By sealing holes on a fly-swatter,we were able to explore the influence of porosity patterning on the aerodynamic coefficients and bistability pendular porous plates. In spite of the simplicity of the considered system, several converging observations allow to draw some general conclusions regarding the role of certain porosity zones. The existence of a sharp stall leads to bistable regimes as a function of the flow velocity in the pendular configuration, with bistability occurring around the stall angle. For a solid square plate, a sharp stall exists and the pendulum displays bistability. No sharp stall has been observed for the other limit case, for which the square is homogeneously porous, leading to a continuous evolution of the pendulum angle with the flow velocity.   
In all tested configurations that present a sharp stall the upper rows are indeed covered and a major part of the holes around the center are also sealed off. Seen in a different light, the bistability of a square plate disappears as soon as holes are opened in the upper rows (i.e. in the immediate vicinity of the leading edge), without impairing the lift production for angles $\theta>\SI{70}{\degree}$. Leading-edge porosity therefore appears as a possibly relevant strategy to dampen stall.
Surface porosity close to the leading-edge on airfoils has also been observed to reduce pressure load due to wing-vortex interactions \citep{Eljack2015}. PIV measurements of the evolution of the wake structure with the angle of attack for the two limit cases (solid and homogenously porous square) clearly demonstrate that the observations made on the global aerodynamic coefficients are linked to the wake characteristics. A detailed study of the influence of the porosity pattern at the immediate vicinity of the leading edge on the wake structure and dynamics, and its relation with the existence of sharp stall was beyond the scope of the present work, but would represent a useful extension. Another aspect  not investigated here is noise reduction induced by the surface porosity. Indeed, porosity at the leading- and trailing-edge is often associated with noise reduction \citep{Zhang2020}. The small vortices induced by the pores destabilizes the large-scale leading- and trailing-edge vortices, which are responsible to a large extent for aircraft noise. This effect of porosity was, in fact, first observed in Nature \citep{Graham1934} and biomimetic concerns spread it to aerospace engineering \citep{Rao2017}. Owls are particularly known for their silent flight and recent studies have shown how the particular structure of their flight feathers enables this feat \citep{Ikeda2018}. The owl feather presents serrations at its leading-edge, and sometimes also throughout the inner vane. Serrations are an ultra-thin comb of barbules and increase the porosity of the feather. The comb breaks the two-dimensionality of the leading-edge vortex which is no longer sustained \citep{Rao2018}. Engineering aerodynamic noise generation by the fine tuning of the surface porosity of objects moving in a flow would also be a possible continuation of this work.



\section*{Acknowledgements}

This work was partly supported by Initiative d'Excellence de
Lyon (IDEXLYON) of the University of Lyon in the
framework of the Programme Investissements d'Avenir
(ANR-16- IDEX-0005) Universit\'e de Lyon. The authors would like to thank Samuel Bera for his involvement in the implementation of the PIV measurements.

\bibliography{article_version_2307}

\begin{thebibliography}{37}%
\makeatletter
\providecommand \@ifxundefined [1]{%
 \@ifx{#1\undefined}
}%
\providecommand \@ifnum [1]{%
 \ifnum #1\expandafter \@firstoftwo
 \else \expandafter \@secondoftwo
 \fi
}%
\providecommand \@ifx [1]{%
 \ifx #1\expandafter \@firstoftwo
 \else \expandafter \@secondoftwo
 \fi
}%
\providecommand \natexlab [1]{#1}%
\providecommand \enquote  [1]{``#1''}%
\providecommand \bibnamefont  [1]{#1}%
\providecommand \bibfnamefont [1]{#1}%
\providecommand \citenamefont [1]{#1}%
\providecommand \href@noop [0]{\@secondoftwo}%
\providecommand \href [0]{\begingroup \@sanitize@url \@href}%
\providecommand \@href[1]{\@@startlink{#1}\@@href}%
\providecommand \@@href[1]{\endgroup#1\@@endlink}%
\providecommand \@sanitize@url [0]{\catcode `\\12\catcode `\$12\catcode `\&12\catcode `\#12\catcode `\^12\catcode `\_12\catcode `\%12\relax}%
\providecommand \@@startlink[1]{}%
\providecommand \@@endlink[0]{}%
\providecommand \url  [0]{\begingroup\@sanitize@url \@url }%
\providecommand \@url [1]{\endgroup\@href {#1}{\urlprefix }}%
\providecommand \urlprefix  [0]{URL }%
\providecommand \Eprint [0]{\href }%
\providecommand \doibase [0]{http://dx.doi.org/}%
\providecommand \selectlanguage [0]{\@gobble}%
\providecommand \bibinfo  [0]{\@secondoftwo}%
\providecommand \bibfield  [0]{\@secondoftwo}%
\providecommand \translation [1]{[#1]}%
\providecommand \BibitemOpen [0]{}%
\providecommand \bibitemStop [0]{}%
\providecommand \bibitemNoStop [0]{.\EOS\space}%
\providecommand \EOS [0]{\spacefactor3000\relax}%
\providecommand \BibitemShut  [1]{\csname bibitem#1\endcsname}%
\let\auto@bib@innerbib\@empty
\bibitem [{\citenamefont {Klebert}\ \emph {et~al.}(2013)\citenamefont {Klebert}, \citenamefont {Lader}, \citenamefont {Gansel},\ and\ \citenamefont {Oppedal}}]{Klebert2013}%
  \BibitemOpen
  \bibfield  {author} {\bibinfo {author} {\bibfnamefont {P.}~\bibnamefont {Klebert}}, \bibinfo {author} {\bibfnamefont {P.}~\bibnamefont {Lader}}, \bibinfo {author} {\bibfnamefont {L.}~\bibnamefont {Gansel}}, \ and\ \bibinfo {author} {\bibfnamefont {F.}~\bibnamefont {Oppedal}},\ }\bibfield  {title} {\enquote {\bibinfo {title} {Hydrodynamic interactions on net panel and aquaculture fish cages: A review},}\ }\href {\doibase https://doi.org/10.1016/j.oceaneng.2012.11.006} {\bibfield  {journal} {\bibinfo  {journal} {Ocean Engineering}\ }\textbf {\bibinfo {volume} {58}},\ \bibinfo {pages} {260--274} (\bibinfo {year} {2013})}\BibitemShut {NoStop}%
\bibitem [{\citenamefont {Laws}\ and\ \citenamefont {Livesey}(1978)}]{Laws1978}%
  \BibitemOpen
  \bibfield  {author} {\bibinfo {author} {\bibfnamefont {E.~M.}\ \bibnamefont {Laws}}\ and\ \bibinfo {author} {\bibfnamefont {J.~L.}\ \bibnamefont {Livesey}},\ }\bibfield  {title} {\enquote {\bibinfo {title} {Flow through screens},}\ }\href {\doibase 10.1146/annurev.fl.10.010178.001335} {\bibfield  {journal} {\bibinfo  {journal} {Annual Review of Fluid Mechanics}\ }\textbf {\bibinfo {volume} {10}},\ \bibinfo {pages} {247--266} (\bibinfo {year} {1978})}\BibitemShut {NoStop}%
\bibitem [{\citenamefont {Lee}\ and\ \citenamefont {Kim}(1999)}]{Lee1999}%
  \BibitemOpen
  \bibfield  {author} {\bibinfo {author} {\bibfnamefont {S.-J.}\ \bibnamefont {Lee}}\ and\ \bibinfo {author} {\bibfnamefont {H.-B.}\ \bibnamefont {Kim}},\ }\bibfield  {title} {\enquote {\bibinfo {title} {Laboratory measurements of velocity and turbulence field behind porous fences},}\ }\href {\doibase https://doi.org/10.1016/S0167-6105(98)00193-7} {\bibfield  {journal} {\bibinfo  {journal} {Journal of Wind Engineering and Industrial Aerodynamics}\ }\textbf {\bibinfo {volume} {80}},\ \bibinfo {pages} {311--326} (\bibinfo {year} {1999})}\BibitemShut {NoStop}%
\bibitem [{\citenamefont {Regalado}\ and\ \citenamefont {Ritter}(2016)}]{Regalado2016}%
  \BibitemOpen
  \bibfield  {author} {\bibinfo {author} {\bibfnamefont {C.~M.}\ \bibnamefont {Regalado}}\ and\ \bibinfo {author} {\bibfnamefont {A.}~\bibnamefont {Ritter}},\ }\bibfield  {title} {\enquote {\bibinfo {title} {The design of an optimal fog water collector: A theoretical analysis},}\ }\href {\doibase https://doi.org/10.1016/j.atmosres.2016.03.006} {\bibfield  {journal} {\bibinfo  {journal} {Atmospheric Research}\ }\textbf {\bibinfo {volume} {178-179}},\ \bibinfo {pages} {45--54} (\bibinfo {year} {2016})}\BibitemShut {NoStop}%
\bibitem [{\citenamefont {Moncuquet}\ \emph {et~al.}(2022)\citenamefont {Moncuquet}, \citenamefont {Mitranescu}, \citenamefont {Marchand}, \citenamefont {Ramananarivo},\ and\ \citenamefont {Duprat}}]{Moncuquet2022}%
  \BibitemOpen
  \bibfield  {author} {\bibinfo {author} {\bibfnamefont {A.}~\bibnamefont {Moncuquet}}, \bibinfo {author} {\bibfnamefont {A.}~\bibnamefont {Mitranescu}}, \bibinfo {author} {\bibfnamefont {O.~C.}\ \bibnamefont {Marchand}}, \bibinfo {author} {\bibfnamefont {S.}~\bibnamefont {Ramananarivo}}, \ and\ \bibinfo {author} {\bibfnamefont {C.}~\bibnamefont {Duprat}},\ }\bibfield  {title} {\enquote {\bibinfo {title} {Collecting fog with vertical fibres: Combined laboratory and in-situ study},}\ }\href {\doibase https://doi.org/10.1016/j.atmosres.2022.106312} {\bibfield  {journal} {\bibinfo  {journal} {Atmospheric Research}\ }\textbf {\bibinfo {volume} {277}},\ \bibinfo {pages} {106312} (\bibinfo {year} {2022})}\BibitemShut {NoStop}%
\bibitem [{\citenamefont {Mittal}, \citenamefont {Ni},\ and\ \citenamefont {Seo}(2020)}]{Mittal2020}%
  \BibitemOpen
  \bibfield  {author} {\bibinfo {author} {\bibfnamefont {R.}~\bibnamefont {Mittal}}, \bibinfo {author} {\bibfnamefont {R.}~\bibnamefont {Ni}}, \ and\ \bibinfo {author} {\bibfnamefont {J.-H.}\ \bibnamefont {Seo}},\ }\bibfield  {title} {\enquote {\bibinfo {title} {The flow physics of covid-19},}\ }\href {\doibase 10.1017/jfm.2020.330} {\bibfield  {journal} {\bibinfo  {journal} {Journal of Fluid Mechanics}\ }\textbf {\bibinfo {volume} {894}},\ \bibinfo {pages} {F2} (\bibinfo {year} {2020})}\BibitemShut {NoStop}%
\bibitem [{\citenamefont {Bourouiba}(2021)}]{Bourouiba2021}%
  \BibitemOpen
  \bibfield  {author} {\bibinfo {author} {\bibfnamefont {L.}~\bibnamefont {Bourouiba}},\ }\bibfield  {title} {\enquote {\bibinfo {title} {The fluid dynamics of disease transmission},}\ }\href {\doibase 10.1146/annurev-fluid-060220-113712} {\bibfield  {journal} {\bibinfo  {journal} {Annual Review of Fluid Mechanics}\ }\textbf {\bibinfo {volume} {53}},\ \bibinfo {pages} {473--508} (\bibinfo {year} {2021})}\BibitemShut {NoStop}%
\bibitem [{\citenamefont {Cummins}\ \emph {et~al.}(2018)\citenamefont {Cummins}, \citenamefont {Seale}, \citenamefont {Macente}, \citenamefont {Certini}, \citenamefont {Mastropaolo}, \citenamefont {Viola},\ and\ \citenamefont {Nakayama}}]{Cummins2018}%
  \BibitemOpen
  \bibfield  {author} {\bibinfo {author} {\bibfnamefont {C.}~\bibnamefont {Cummins}}, \bibinfo {author} {\bibfnamefont {M.}~\bibnamefont {Seale}}, \bibinfo {author} {\bibfnamefont {A.}~\bibnamefont {Macente}}, \bibinfo {author} {\bibfnamefont {D.}~\bibnamefont {Certini}}, \bibinfo {author} {\bibfnamefont {E.}~\bibnamefont {Mastropaolo}}, \bibinfo {author} {\bibfnamefont {I.~M.}\ \bibnamefont {Viola}}, \ and\ \bibinfo {author} {\bibfnamefont {N.}~\bibnamefont {Nakayama}},\ }\bibfield  {title} {\enquote {\bibinfo {title} {A separated vortex ring underlies the flight of the dandelion},}\ }\href {\doibase 10.1038/s41586-018-0604-2} {\bibfield  {journal} {\bibinfo  {journal} {Nature}\ }\textbf {\bibinfo {volume} {562}},\ \bibinfo {pages} {414--418} (\bibinfo {year} {2018})}\BibitemShut {NoStop}%
\bibitem [{\citenamefont {Ledda}\ \emph {et~al.}(2019)\citenamefont {Ledda}, \citenamefont {Siconolfi}, \citenamefont {Viola}, \citenamefont {Camarri},\ and\ \citenamefont {Gallaire}}]{Ledda2019}%
  \BibitemOpen
  \bibfield  {author} {\bibinfo {author} {\bibfnamefont {P.~G.}\ \bibnamefont {Ledda}}, \bibinfo {author} {\bibfnamefont {L.}~\bibnamefont {Siconolfi}}, \bibinfo {author} {\bibfnamefont {F.}~\bibnamefont {Viola}}, \bibinfo {author} {\bibfnamefont {S.}~\bibnamefont {Camarri}}, \ and\ \bibinfo {author} {\bibfnamefont {F.}~\bibnamefont {Gallaire}},\ }\bibfield  {title} {\enquote {\bibinfo {title} {Flow dynamics of a dandelion pappus: A linear stability approach},}\ }\href {\doibase 10.1103/PhysRevFluids.4.071901} {\bibfield  {journal} {\bibinfo  {journal} {Phys. Rev. Fluids}\ }\textbf {\bibinfo {volume} {4}},\ \bibinfo {pages} {071901(R)} (\bibinfo {year} {2019})}\BibitemShut {NoStop}%
\bibitem [{\citenamefont {Sunada}\ \emph {et~al.}(2002)\citenamefont {Sunada}, \citenamefont {Takashima}, \citenamefont {Hattori}, \citenamefont {Yasuda},\ and\ \citenamefont {Kawachi}}]{Sunada2002}%
  \BibitemOpen
  \bibfield  {author} {\bibinfo {author} {\bibfnamefont {S.}~\bibnamefont {Sunada}}, \bibinfo {author} {\bibfnamefont {H.}~\bibnamefont {Takashima}}, \bibinfo {author} {\bibfnamefont {T.}~\bibnamefont {Hattori}}, \bibinfo {author} {\bibfnamefont {K.}~\bibnamefont {Yasuda}}, \ and\ \bibinfo {author} {\bibfnamefont {K.}~\bibnamefont {Kawachi}},\ }\bibfield  {title} {\enquote {\bibinfo {title} {{Fluid-dynamic characteristics of a bristled wing}},}\ }\href {\doibase 10.1242/jeb.205.17.2737} {\bibfield  {journal} {\bibinfo  {journal} {Journal of Experimental Biology}\ }\textbf {\bibinfo {volume} {205}},\ \bibinfo {pages} {2737--2744} (\bibinfo {year} {2002})}\BibitemShut {NoStop}%
\bibitem [{\citenamefont {Kolomenskiy}\ \emph {et~al.}(2020)\citenamefont {Kolomenskiy}, \citenamefont {Farisenkov}, \citenamefont {Engels}, \citenamefont {Lapina}, \citenamefont {Petrov}, \citenamefont {Lehmann}, \citenamefont {Onishi}, \citenamefont {Liu},\ and\ \citenamefont {Polilov}}]{Kolomenskiy2020}%
  \BibitemOpen
  \bibfield  {author} {\bibinfo {author} {\bibfnamefont {D.}~\bibnamefont {Kolomenskiy}}, \bibinfo {author} {\bibfnamefont {S.}~\bibnamefont {Farisenkov}}, \bibinfo {author} {\bibfnamefont {T.}~\bibnamefont {Engels}}, \bibinfo {author} {\bibfnamefont {N.}~\bibnamefont {Lapina}}, \bibinfo {author} {\bibfnamefont {P.}~\bibnamefont {Petrov}}, \bibinfo {author} {\bibfnamefont {F.~O.}\ \bibnamefont {Lehmann}}, \bibinfo {author} {\bibfnamefont {R.}~\bibnamefont {Onishi}}, \bibinfo {author} {\bibfnamefont {H.}~\bibnamefont {Liu}}, \ and\ \bibinfo {author} {\bibfnamefont {A.}~\bibnamefont {Polilov}},\ }\bibfield  {title} {\enquote {\bibinfo {title} {{Aerodynamic performance of a bristled wing of a very small insect: Dynamically scaled model experiments and computational fluid dynamics simulations using a revolving wing model}},}\ }\href {\doibase 10.1007/s00348-020-03027-0} {\bibfield  {journal} {\bibinfo  {journal} {Experiments in Fluids}\ }\textbf {\bibinfo {volume} {61}},\ \bibinfo {pages} {1--13} (\bibinfo {year}
  {2020})}\BibitemShut {NoStop}%
\bibitem [{\citenamefont {Taylor}(1944)}]{Taylor1944_1}%
  \BibitemOpen
  \bibfield  {author} {\bibinfo {author} {\bibfnamefont {G.~I.}\ \bibnamefont {Taylor}},\ }\href@noop {} {\enquote {\bibinfo {title} {Air resistance of a flat plate of very porous material},}\ }\bibinfo {type} {Tech. Rep.}\ \bibinfo {number} {2236}\ (\bibinfo  {institution} {Ministry of Supply, Aeronautical Research Council},\ \bibinfo {address} {London {$[$}England{$]$}},\ \bibinfo {year} {1944})\BibitemShut {NoStop}%
\bibitem [{\citenamefont {Taylor}\ and\ \citenamefont {Davies}(1944)}]{Taylor1944_2}%
  \BibitemOpen
  \bibfield  {author} {\bibinfo {author} {\bibfnamefont {G.~I.}\ \bibnamefont {Taylor}}\ and\ \bibinfo {author} {\bibfnamefont {R.~M.}\ \bibnamefont {Davies}},\ }\href {\doibase https://worldcat.org/title/842094201} {\enquote {\bibinfo {title} {The aerodynamics of porous sheets},}\ }\bibinfo {type} {Tech. Rep.}\ \bibinfo {number} {2237}\ (\bibinfo  {institution} {Ministry of Supply, Aeronautical Research Council},\ \bibinfo {year} {1944})\BibitemShut {NoStop}%
\bibitem [{\citenamefont {Villermaux}\ and\ \citenamefont {Hopfinger}(1994)}]{Villermaux1994}%
  \BibitemOpen
  \bibfield  {author} {\bibinfo {author} {\bibfnamefont {E.}~\bibnamefont {Villermaux}}\ and\ \bibinfo {author} {\bibfnamefont {E.~J.}\ \bibnamefont {Hopfinger}},\ }\bibfield  {title} {\enquote {\bibinfo {title} {Periodically arranged co-flowing jets},}\ }\href {\doibase 10.1017/S0022112094004039} {\bibfield  {journal} {\bibinfo  {journal} {Journal of Fluid Mechanics}\ }\textbf {\bibinfo {volume} {263}},\ \bibinfo {pages} {63--92} (\bibinfo {year} {1994})}\BibitemShut {NoStop}%
\bibitem [{\citenamefont {Steiros}\ \emph {et~al.}(2020)\citenamefont {Steiros}, \citenamefont {Kokmanian}, \citenamefont {Bempedelis},\ and\ \citenamefont {Hultmark}}]{Steiros2020}%
  \BibitemOpen
  \bibfield  {author} {\bibinfo {author} {\bibfnamefont {K.}~\bibnamefont {Steiros}}, \bibinfo {author} {\bibfnamefont {K.}~\bibnamefont {Kokmanian}}, \bibinfo {author} {\bibfnamefont {N.}~\bibnamefont {Bempedelis}}, \ and\ \bibinfo {author} {\bibfnamefont {M.}~\bibnamefont {Hultmark}},\ }\bibfield  {title} {\enquote {\bibinfo {title} {The effect of porosity on the drag of cylinders},}\ }\href {\doibase 10.1017/jfm.2020.606} {\bibfield  {journal} {\bibinfo  {journal} {Journal of Fluid Mechanics}\ }\textbf {\bibinfo {volume} {901}},\ \bibinfo {pages} {R2} (\bibinfo {year} {2020})}\BibitemShut {NoStop}%
\bibitem [{\citenamefont {Koo}\ and\ \citenamefont {James}(1973)}]{Koo1973}%
  \BibitemOpen
  \bibfield  {author} {\bibinfo {author} {\bibfnamefont {J.-K.}\ \bibnamefont {Koo}}\ and\ \bibinfo {author} {\bibfnamefont {D.~F.}\ \bibnamefont {James}},\ }\bibfield  {title} {\enquote {\bibinfo {title} {Fluid flow around and through a screen},}\ }\href {\doibase 10.1017/S0022112073000327} {\bibfield  {journal} {\bibinfo  {journal} {Journal of Fluid Mechanics}\ }\textbf {\bibinfo {volume} {60}},\ \bibinfo {pages} {513--538} (\bibinfo {year} {1973})}\BibitemShut {NoStop}%
\bibitem [{\citenamefont {Steiros}\ and\ \citenamefont {Hultmark}(2018)}]{Steiros2018}%
  \BibitemOpen
  \bibfield  {author} {\bibinfo {author} {\bibfnamefont {K.}~\bibnamefont {Steiros}}\ and\ \bibinfo {author} {\bibfnamefont {M.}~\bibnamefont {Hultmark}},\ }\bibfield  {title} {\enquote {\bibinfo {title} {Drag on flat plates of arbitrary porosity},}\ }\href {\doibase 10.1017/jfm.2018.621} {\bibfield  {journal} {\bibinfo  {journal} {Journal of Fluid Mechanics}\ }\textbf {\bibinfo {volume} {853}},\ \bibinfo {pages} {R3} (\bibinfo {year} {2018})}\BibitemShut {NoStop}%
\bibitem [{\citenamefont {Marchand}\ \emph {et~al.}(2023)\citenamefont {Marchand}, \citenamefont {Ramananarivo}, \citenamefont {Duprat},\ and\ \citenamefont {Josserand}}]{Marchand2023}%
  \BibitemOpen
  \bibfield  {author} {\bibinfo {author} {\bibfnamefont {O.~C.}\ \bibnamefont {Marchand}}, \bibinfo {author} {\bibfnamefont {S.}~\bibnamefont {Ramananarivo}}, \bibinfo {author} {\bibfnamefont {C.}~\bibnamefont {Duprat}}, \ and\ \bibinfo {author} {\bibfnamefont {C.}~\bibnamefont {Josserand}},\ }\href@noop {} {\enquote {\bibinfo {title} {Three-dimensional flow around and through a porous screen},}\ } (\bibinfo {year} {2023}),\ \bibinfo {note} {arXiv:2303.00711}\BibitemShut {NoStop}%
\bibitem [{\citenamefont {Vincent}, \citenamefont {Shambaugh},\ and\ \citenamefont {Kanso}(2016)}]{Vincent2016}%
  \BibitemOpen
  \bibfield  {author} {\bibinfo {author} {\bibfnamefont {L.}~\bibnamefont {Vincent}}, \bibinfo {author} {\bibfnamefont {W.~S.}\ \bibnamefont {Shambaugh}}, \ and\ \bibinfo {author} {\bibfnamefont {E.}~\bibnamefont {Kanso}},\ }\bibfield  {title} {\enquote {\bibinfo {title} {{Holes stabilize freely falling coins}},}\ }\href {\doibase 10.1017/jfm.2016.432} {\bibfield  {journal} {\bibinfo  {journal} {Journal of Fluid Mechanics}\ }\textbf {\bibinfo {volume} {801}},\ \bibinfo {pages} {250--259} (\bibinfo {year} {2016})}\BibitemShut {NoStop}%
\bibitem [{\citenamefont {Vagnoli}\ \emph {et~al.}(2023)\citenamefont {Vagnoli}, \citenamefont {Zampogna}, \citenamefont {Camarri}, \citenamefont {Gallaire},\ and\ \citenamefont {Ledda}}]{Vagnoli2022}%
  \BibitemOpen
  \bibfield  {author} {\bibinfo {author} {\bibfnamefont {G.}~\bibnamefont {Vagnoli}}, \bibinfo {author} {\bibfnamefont {G.}~\bibnamefont {Zampogna}}, \bibinfo {author} {\bibfnamefont {S.}~\bibnamefont {Camarri}}, \bibinfo {author} {\bibfnamefont {F.}~\bibnamefont {Gallaire}}, \ and\ \bibinfo {author} {\bibfnamefont {P.}~\bibnamefont {Ledda}},\ }\bibfield  {title} {\enquote {\bibinfo {title} {Permeability sets the linear path instability of buoyancy-driven disks},}\ }\href {\doibase 10.1017/jfm.2022.989} {\bibfield  {journal} {\bibinfo  {journal} {Journal of Fluid Mechanics}\ }\textbf {\bibinfo {volume} {955}},\ \bibinfo {pages} {A29} (\bibinfo {year} {2023})}\BibitemShut {NoStop}%
\bibitem [{\citenamefont {Montgomery}(1900)}]{Montgomery1900}%
  \BibitemOpen
  \bibfield  {author} {\bibinfo {author} {\bibfnamefont {R.~R.}\ \bibnamefont {Montgomery}},\ }\href@noop {} {\enquote {\bibinfo {title} {{Fly killer}},}\ } (\bibinfo {year} {1900})\BibitemShut {NoStop}%
\bibitem [{\citenamefont {Gatch}(1927)}]{Gatch1927}%
  \BibitemOpen
  \bibfield  {author} {\bibinfo {author} {\bibfnamefont {M.~W.}\ \bibnamefont {Gatch}},\ }\href@noop {} {\enquote {\bibinfo {title} {{Fly swatter}},}\ } (\bibinfo {year} {1927})\BibitemShut {NoStop}%
\bibitem [{\citenamefont {Brownson}(1938)}]{Brownson1938}%
  \BibitemOpen
  \bibfield  {author} {\bibinfo {author} {\bibfnamefont {P.~J.}\ \bibnamefont {Brownson}},\ }\href@noop {} {\enquote {\bibinfo {title} {{Fly swatter}},}\ } (\bibinfo {year} {1938})\BibitemShut {NoStop}%
\bibitem [{\citenamefont {Baker}(1939)}]{Baker1939}%
  \BibitemOpen
  \bibfield  {author} {\bibinfo {author} {\bibfnamefont {D.}~\bibnamefont {Baker}},\ }\href@noop {} {\enquote {\bibinfo {title} {{Fly swatter}},}\ } (\bibinfo {year} {1939})\BibitemShut {NoStop}%
\bibitem [{\citenamefont {Obligado}, \citenamefont {Puy},\ and\ \citenamefont {Bourgoin}(2013)}]{Obligado2013}%
  \BibitemOpen
  \bibfield  {author} {\bibinfo {author} {\bibfnamefont {M.}~\bibnamefont {Obligado}}, \bibinfo {author} {\bibfnamefont {M.}~\bibnamefont {Puy}}, \ and\ \bibinfo {author} {\bibfnamefont {M.}~\bibnamefont {Bourgoin}},\ }\bibfield  {title} {\enquote {\bibinfo {title} {Bi-stability of a pendular disk in laminar and turbulent flows},}\ }\href {\doibase 10.1017/jfm.2013.312} {\bibfield  {journal} {\bibinfo  {journal} {Journal of Fluid Mechanics}\ }\textbf {\bibinfo {volume} {728}},\ \bibinfo {pages} {R2} (\bibinfo {year} {2013})}\BibitemShut {NoStop}%
\bibitem [{\citenamefont {Gayout}, \citenamefont {Bourgoin},\ and\ \citenamefont {Plihon}(2021)}]{Gayout2021}%
  \BibitemOpen
  \bibfield  {author} {\bibinfo {author} {\bibfnamefont {A.}~\bibnamefont {Gayout}}, \bibinfo {author} {\bibfnamefont {M.}~\bibnamefont {Bourgoin}}, \ and\ \bibinfo {author} {\bibfnamefont {N.}~\bibnamefont {Plihon}},\ }\bibfield  {title} {\enquote {\bibinfo {title} {Rare event-triggered transitions in aerodynamic bifurcation},}\ }\href {\doibase 10.1103/PhysRevLett.126.104501} {\bibfield  {journal} {\bibinfo  {journal} {Phys. Rev. Lett.}\ }\textbf {\bibinfo {volume} {126}},\ \bibinfo {pages} {104501} (\bibinfo {year} {2021})}\BibitemShut {NoStop}%
\bibitem [{\citenamefont {Gayout}\ \emph {et~al.}(2023)\citenamefont {Gayout}, \citenamefont {Gylfason}, \citenamefont {Plihon},\ and\ \citenamefont {Bourgoin}}]{Gayout2023}%
  \BibitemOpen
  \bibfield  {author} {\bibinfo {author} {\bibfnamefont {A.}~\bibnamefont {Gayout}}, \bibinfo {author} {\bibfnamefont {A.}~\bibnamefont {Gylfason}}, \bibinfo {author} {\bibfnamefont {N.}~\bibnamefont {Plihon}}, \ and\ \bibinfo {author} {\bibfnamefont {M.}~\bibnamefont {Bourgoin}},\ }\bibfield  {title} {\enquote {\bibinfo {title} {Fluidelastic modeling of a weathercock stabilization in a uniform flow},}\ }\href {\doibase https://doi.org/10.1016/j.jfluidstructs.2023.103895} {\bibfield  {journal} {\bibinfo  {journal} {Journal of Fluids and Structures}\ }\textbf {\bibinfo {volume} {120}},\ \bibinfo {pages} {103895} (\bibinfo {year} {2023})}\BibitemShut {NoStop}%
\bibitem [{\citenamefont {Flachsbart}(1932)}]{Flachsbart1932}%
  \BibitemOpen
  \bibfield  {author} {\bibinfo {author} {\bibfnamefont {O.}~\bibnamefont {Flachsbart}},\ }\bibfield  {title} {\enquote {\bibinfo {title} {{Messungen an ebenen und gew{\"{o}}lbten Platten}},}\ }in\ \href {\doibase https://doi.org/10.1515/9783486764406-017} {\emph {\bibinfo {booktitle} {Ergebnisse der Aerodynamischen Versuchsanstalt zu G{\"{o}}ttingen - IV. Lieferung}}}\ (\bibinfo  {publisher} {Verlag von R. Oldenburg},\ \bibinfo {address} {M{\"{u}}nchen und Berlin},\ \bibinfo {year} {1932})\ pp.\ \bibinfo {pages} {96--100}\BibitemShut {NoStop}%
\bibitem [{\citenamefont {Gao}\ \emph {et~al.}(2018)\citenamefont {Gao}, \citenamefont {Tao}, \citenamefont {Tian},\ and\ \citenamefont {Yang}}]{Gao2018}%
  \BibitemOpen
  \bibfield  {author} {\bibinfo {author} {\bibfnamefont {S.}~\bibnamefont {Gao}}, \bibinfo {author} {\bibfnamefont {L.}~\bibnamefont {Tao}}, \bibinfo {author} {\bibfnamefont {X.}~\bibnamefont {Tian}}, \ and\ \bibinfo {author} {\bibfnamefont {J.}~\bibnamefont {Yang}},\ }\bibfield  {title} {\enquote {\bibinfo {title} {{Flow around an inclined circular disk}},}\ }\href {\doibase 10.1017/jfm.2018.526} {\bibfield  {journal} {\bibinfo  {journal} {J. Fluid Mech.}\ }\textbf {\bibinfo {volume} {851}},\ \bibinfo {pages} {687--714} (\bibinfo {year} {2018})}\BibitemShut {NoStop}%
\bibitem [{\citenamefont {Sommeria}(2008)}]{bib:uvmat}%
  \BibitemOpen
  \bibfield  {author} {\bibinfo {author} {\bibfnamefont {J.}~\bibnamefont {Sommeria}},\ }\bibfield  {title} {\enquote {\bibinfo {title} {{UVMAT}},}\ }\href {http://servforge.legi.grenoble-inp.fr/projects/soft-uvmat} {\bibfield  {journal} {\bibinfo  {journal} {http://servforge.legi.grenoble-inp.fr/projects/soft-uvmat}\ } (\bibinfo {year} {2008})}\BibitemShut {NoStop}%
\bibitem [{\citenamefont {Eiffel}(1910)}]{Eiffel1910}%
  \BibitemOpen
  \bibfield  {author} {\bibinfo {author} {\bibfnamefont {G.}~\bibnamefont {Eiffel}},\ }\bibfield  {title} {\enquote {\bibinfo {title} {{Sur la r{\'{e}}sistance des plans rectangulaires frapp{\'{e}}s obliquement par le vent}},}\ }\href {https://gallica.bnf.fr/ark:/12148/bpt6k31042.item} {\bibfield  {journal} {\bibinfo  {journal} {Comptes-rendus hebdomadaires des s{\'{e}}ances de l'Acad{\'{e}}mie des Sciences}\ }\textbf {\bibinfo {volume} {151}},\ \bibinfo {pages} {979--981} (\bibinfo {year} {1910})}\BibitemShut {NoStop}%
\bibitem [{\citenamefont {Eljack}, \citenamefont {AlQadi},\ and\ \citenamefont {Khalid}(2015)}]{Eljack2015}%
  \BibitemOpen
  \bibfield  {author} {\bibinfo {author} {\bibfnamefont {E.}~\bibnamefont {Eljack}}, \bibinfo {author} {\bibfnamefont {I.}~\bibnamefont {AlQadi}}, \ and\ \bibinfo {author} {\bibfnamefont {M.}~\bibnamefont {Khalid}},\ }\bibfield  {title} {\enquote {\bibinfo {title} {{Numerical simulation of surface porosity in presence of wing-vortex interaction}},}\ }\href {\doibase 10.1108/AEAT-09-2013-0165} {\bibfield  {journal} {\bibinfo  {journal} {Aircraft Engineering and Aerospace Technology}\ }\textbf {\bibinfo {volume} {87}},\ \bibinfo {pages} {443--453} (\bibinfo {year} {2015})}\BibitemShut {NoStop}%
\bibitem [{\citenamefont {Zhang}\ and\ \citenamefont {Chong}(2020)}]{Zhang2020}%
  \BibitemOpen
  \bibfield  {author} {\bibinfo {author} {\bibfnamefont {M.}~\bibnamefont {Zhang}}\ and\ \bibinfo {author} {\bibfnamefont {T.~P.}\ \bibnamefont {Chong}},\ }\bibfield  {title} {\enquote {\bibinfo {title} {{Experimental investigation of the impact of porous parameters on trailing-edge noise}},}\ }\href {\doibase 10.1016/j.jsv.2020.115694} {\bibfield  {journal} {\bibinfo  {journal} {Journal of Sound and Vibration}\ }\textbf {\bibinfo {volume} {489}},\ \bibinfo {pages} {115694} (\bibinfo {year} {2020})}\BibitemShut {NoStop}%
\bibitem [{\citenamefont {Graham}(1934)}]{Graham1934}%
  \BibitemOpen
  \bibfield  {author} {\bibinfo {author} {\bibfnamefont {R.~R.}\ \bibnamefont {Graham}},\ }\bibfield  {title} {\enquote {\bibinfo {title} {{The Silent Flight of Owls}},}\ }\href {\doibase 10.1017/S0368393100109915} {\bibfield  {journal} {\bibinfo  {journal} {The Journal of the Royal Aeronautical Society}\ }\textbf {\bibinfo {volume} {38}},\ \bibinfo {pages} {837--843} (\bibinfo {year} {1934})}\BibitemShut {NoStop}%
\bibitem [{\citenamefont {Rao}\ \emph {et~al.}(2017)\citenamefont {Rao}, \citenamefont {Ikeda}, \citenamefont {Nakata},\ and\ \citenamefont {Liu}}]{Rao2017}%
  \BibitemOpen
  \bibfield  {author} {\bibinfo {author} {\bibfnamefont {C.}~\bibnamefont {Rao}}, \bibinfo {author} {\bibfnamefont {T.}~\bibnamefont {Ikeda}}, \bibinfo {author} {\bibfnamefont {T.}~\bibnamefont {Nakata}}, \ and\ \bibinfo {author} {\bibfnamefont {H.}~\bibnamefont {Liu}},\ }\bibfield  {title} {\enquote {\bibinfo {title} {{Owl-inspired leading-edge serrations play a crucial role in aerodynamic force production and sound suppression}},}\ }\href {\doibase 10.1088/1748-3190/aa7013} {\bibfield  {journal} {\bibinfo  {journal} {Bioinspiration {\&} Biomimetics}\ }\textbf {\bibinfo {volume} {12}},\ \bibinfo {pages} {046008} (\bibinfo {year} {2017})}\BibitemShut {NoStop}%
\bibitem [{\citenamefont {Ikeda}\ \emph {et~al.}(2018)\citenamefont {Ikeda}, \citenamefont {Ueda}, \citenamefont {Nakata}, \citenamefont {Noda}, \citenamefont {Tanaka}, \citenamefont {Fujii},\ and\ \citenamefont {Liu}}]{Ikeda2018}%
  \BibitemOpen
  \bibfield  {author} {\bibinfo {author} {\bibfnamefont {T.}~\bibnamefont {Ikeda}}, \bibinfo {author} {\bibfnamefont {T.}~\bibnamefont {Ueda}}, \bibinfo {author} {\bibfnamefont {T.}~\bibnamefont {Nakata}}, \bibinfo {author} {\bibfnamefont {R.}~\bibnamefont {Noda}}, \bibinfo {author} {\bibfnamefont {H.}~\bibnamefont {Tanaka}}, \bibinfo {author} {\bibfnamefont {T.}~\bibnamefont {Fujii}}, \ and\ \bibinfo {author} {\bibfnamefont {H.}~\bibnamefont {Liu}},\ }\bibfield  {title} {\enquote {\bibinfo {title} {{Morphology Effects of Leading-edge Serrations on Aerodynamic Force Production: An Integrated Study Using PIV and Force Measurements}},}\ }\href {\doibase 10.1007/s42235-018-0054-4} {\bibfield  {journal} {\bibinfo  {journal} {Journal of Bionic Engineering}\ }\textbf {\bibinfo {volume} {15}},\ \bibinfo {pages} {661--672} (\bibinfo {year} {2018})}\BibitemShut {NoStop}%
\bibitem [{\citenamefont {Rao}\ and\ \citenamefont {Liu}(2018)}]{Rao2018}%
  \BibitemOpen
  \bibfield  {author} {\bibinfo {author} {\bibfnamefont {C.}~\bibnamefont {Rao}}\ and\ \bibinfo {author} {\bibfnamefont {H.}~\bibnamefont {Liu}},\ }\bibfield  {title} {\enquote {\bibinfo {title} {{Aerodynamic robustness in owl-inspired leading-edge serrations: a computational wind-gust model}},}\ }\href {\doibase 10.1088/1748-3190/aacb43} {\bibfield  {journal} {\bibinfo  {journal} {Bioinspiration {\&} Biomimetics}\ }\textbf {\bibinfo {volume} {13}},\ \bibinfo {pages} {056002} (\bibinfo {year} {2018})}\BibitemShut {NoStop}%
\end{thebibliography}%

\end{document}